# Expectation-Maximization (EM) Algorithms for Mapping Short Reads Illustrated with FAIRE data and the TP53-WRAP53 Gene Region

Peter J. Waddell[1] and Timothy Herston[2]

pwaddell@purdue.edu

[1]Department of Biological Sciences, Purdue University, West Lafayette, IN 47906, U.S.A.
[2]School of Liberal Arts and Sciences, Ivy Tech Community College, Lafayette, IN 47906, U.S.A

Huge numbers of short reads are being generated for mapping back to the genome to discover the frequency of transcripts, miRNAs, DNAase hypersensitive sites, FAIRE regions, nucleosome occupancy, etc. Since these reads are typically short (e.g., 36 base pairs) and since many eukaryotic genomes, including humans, have highly repetitive sequences then many of these reads map to two or more locations in the genome. Current mapping of these reads, grading them according to 0, 1 or 2 mismatches wastes a great deal of information. These short sequences are typically mapped with no account of the accuracy of the sequence, even in company software when per base error rates are being reported by another part of the machine.  Further, multiply mapping locations are frequently discarded altogether or allocated with no regard to where other reads are accumulating. Here we show how to combine probabilistic mapping of reads with an EM algorithm to iteratively improve the empirical likelihood of the allocation of short reads. Mapping using LAST takes into account the per base accuracy of the read, plus insertions and deletions, plus anticipated occasional errors or SNPs with respect to the parent genome. The probabilistic EM algorithm iteratively allocates reads based on the proportion of reads mapping within windows on the previous cycle, along with any prior information on where the read best maps. The methods are illustrated with FAIRE ENCODE data looking at the very important head-to-head gene combination of TP53 and WRAP 53.

# 1 Introduction

Next generation or massively parallel sequencing has made possible the use of short reads from randomly fragmented DNA useful in a number of genomic contexts. Assuming the reads are truly randomly distributed with respect to the source DNA, they can be used to not only reconstruct the original DNA or RNA sequence, but to infer the relative abundance of the original sequences. That is, they can be used to estimate the relative abundance of miRNAs, mRNA transcripts, protein-bound DNA regions, or open DNA, such as FAIRE regions (e.g., Bhinge et al. 2007, Johnson et al. 2007, Mortazavi et al. 2008, Wang et al. 2007, 2008, ENOCDE Project Consortium 2007, Giresi et al. 2007). This is done by matching (mapping) the reads to the source (target) sequence database (e.g., the collection of all mRNA's or the genome, depending on the exact use) and then calculating the relative frequency of matches at different genomic locations.

Since most genomes contain repetitive sequences, then there are often short reads that match equally well to two or more places in the genome. Such reads are called ambiguously mapping reads, or just ambiguous reads. Mammalian genomes are a good example of this with a large fraction of the genome originating as L1 retrotransposable elements (e.g., Waters et al. 2007), and these generate many nearly equally good locations to map a read to. In our experience, greater than 50% of 36 base pair reads from the Illumina Solexa sequencing system can map to more than one location in a mammalian genome with the same minimal mismatch score.

Making the situation more difficult, the latest massively parallel sequencing technologies tend to have per base error rates far higher than the previous generation of machines (e.g., the ABI 3730). Error rates of 1 to 5 per 100 bases seem common in practice, either across the whole sequence or at specific positions. While such error rates are typically swamped when massively over-sequencing a region (e.g., Harismendy et al. 2009), they remain a serious issue when mapping short reads back to the genome for the purpose of assessing the frequency of different types of RNA and DNA. Typically, the error rates tend to rise towards the end of the read (while the first few bases can also suffer technical problems). Fortunately, the machines can also output per base error rate estimates. The 454 machine is notorious for misreading the lengths of homopolymers and will often call a five base stretch of A, for example, as a four or a six base stretch. In contrast, the Illumina Solexa machine, because of the chemistry used, appears to very rarely call either extra bases, nor does it miss calling a base (e.g., Harismendy et al. 2009).

Present day software, for example the Illumina ELAND software (Illumina 2009, www.illumina.com), maps reads based only on minimum mismatch score (with up to two mismatches allowed). It makes no distinction as to the type of error (e.g., A goes to G versus A goes to C), per base error estimates, or the possibility of an indel (DNA base insertion or deletion) in either the machine's read, the reference genome or the DNA sample being analyzed. Mapping only unambiguous locations is going to under represent any sequence with repetitive structure, and in some cases (such as mapping transcription start sites) might result in completely missing a site and along with it, the functional role of repetitive motifs. It would also be highly desirable to incorporate all relevant mapping error probabilities in a meaningful way. These probabilities, in reality, frequently differ by multiples or even orders of magnitude from each other, so there is a lot of potentially useful information being discarded by mapping using just mismatches.

The treatment of ambiguous reads by ELAND and other currently available software is quite problematic. The basic mode of operation is that if a read has just one location with the minimal mismatch score being 0, 1 or 2, it is kept and mapped, else it is discarded. Apart from throwing away information such as discarding a perfect match except for the last 3 bases were sequencing has gone awry, this introduces a clear bias against mapping reads to repetitive parts of the genome.

A refinement is to distribute ambiguous reads in proportion to the number of locations they map to (e.g., Li et al. 2008, in the popular package SOAP). For example, if a read maps with minimal mismatch score to one to three locations in the genome, each location is given a partial weight of 1/3. This is used by a number of packages currently. It uses more of the information, but produces biases. If a read could come from two regions of the genome, but one already has support of 100 unambiguous reads and the other only 2 reads, then it would seem more intuitively reasonable that the read came from the former region. All things being equal and without any other information, the distribution ratio might then be 100/102 to the former and only 2/102 to the latter region. If the former case had all 102 prior reads in one location and none in the other, the call might be 102/102 and 0/102.

While the above proportional procedure is an improvement, it can be improved upon. Once all regions have been mapped as in the paragraph above, then the whole procedure can start again, except that the previous support for regions are the proportionality weights for ambiguous reads. This is basically what is called an Expectation-Maximization or an EM algorithm (Dempster et al., 1977). In this case, while 102/102 and 0/102 may be maximum likelihood estimates of relative support, it may be desirable to "lift" such cases of the sticky boundary of zero. In this case, the first cycle might give apportion 101.5/102 and 0.5/102, and then let the subsequent cycles decide if the 0.5 is going to zero, or increasing, perhaps due to it being the source of other ambiguous reads. An example of an expectation-maximization or EM algorithm being used to resolve another type of DNA sequence data with ambiguous information is the calculation of the unconstrained likelihood of a set of aligned sequences (Waddell, 2005).

In our own case, we are particularly interested in mapping FAIRE (Formaldehyde Assisted Identification of Regulatory Elements) derived DNA (Giresi et al. 2007) back to the genome for comparative studies (e.g., Wray 2003). This is slightly different to mapping reads to transcripts, in that there are no predetermined boundaries to windows and a read might span the boundary of an arbitrary "window". FAIRE DNA itself appears to be predominantly nucleosome free or "open" DNA. This DNA is isolated by cross linking DNA to protein with formaldehyde, sonicating DNA to about the size of a nucleosome (100-200 base pairs), extracting the 1-3% of total DNA that does not have protein stuck on it (the phenol phase of a phenol/chloroform DNA extraction), repairing the blunt ends, ligating on adaptors (possibly with PCR to increase abundance and select for competent product) then mass sequencing millions of these molecules with 36 base pair or longer reads from one end (or more rarely, both ends).

To detect the open regions it is necessary to map reads from the FAIRE procedure to fairly narrow windows (50 to 200 bases seems to work well) in order to give some resolution as to the exact location of particularly open regions. Since the boundaries of these regions are not known *a priori* (and may indeed be dynamic within a cell population), it is possible to either set non-overlapping windows or map with overlapping windows. If sliding, the amount to slide is ideally 1 base pair (bp) at a time, but sliding 10bp does little to reduce the desired resolution and reduces computational overhead. The window used to assess the support for an ambiguous read is the window upon which it is closest to centered. This can create some ambiguities when reads map to adjacent overlapping windows. This seems to be relatively minor due to the small size of the windows being used. Its impact can be assessed by rerunning the algorithm with the slide set to the same size as the window (hence zero overlap). Since reads can fall across the boundaries of windows (overlapping or not), it seems preferable to allocate just the fraction that falls within the window, rather than a more arbitrary all or nothing.

A second factor is that mismatch scores are not necessarily ideal for mapping short reads for these types of purpose. Far more desirable is to map the probability of the read given the sequencing machines error file, while simultaneously allowing for Simple Nucleotide Polymorphisms (SNPs) such as single base substitutions and short indels (Frith et al. 2009). Information on the former probabilities can be found in the *.PRB files output by the Illumina Solexa pipeline, with a probability of misreading the base in all four possible ways specific to every position in the read (thus there are $4 \times n$ probabilities for a read *n* bases long). A simplified and potentially less informative version is found in the FASTQ output file, where the probability of any type of misread is given for each position (analogous to the PHRED scores of the older generation capillary machines).

Estimates of SNP rates in the genome being studied may be available from the genome assembly. These rates can vary widely amongst species. Even for highly homozygous species, it may be advisable not to decrease the single nucleotide polymorphism probability too much lower than 1/2000 to guard against overly optimistic scores in the PRB files (Frith et al. 2009) and the fact that there are parts of most genomes that do seem to have markedly higher SNP rates than average.

One of the best studied regions of the human genome is the TP53 gene region. Transcription of TP53 is the first line of regulation in this very important gene. A second line of regulation recently shown to be mediated by TP53's head to head partner (a gene of many names including WRAP53, WD40, GNB5-like) is stabilization of the TP53 transcript alone by presence of a rare extra-long anti-sense WRAP53 transcript (even though it is at a much lower stoichiometry and covers only the short first exon of the major TP53 transcript, Mahmoudi et al. 2009). Since WRAP53 also appears to be essential for the activity of the telomerase complex (Venteicher et al. 2009), this appears to be a critical part of the regulation of TP53, with the potential aim of preventing the "loose cannon" TP53 going off and killing the cell when telomerase is inactive, and hence the cell cannot be in a malignant state. Whether such an on-off signal is overridden by interferon signaling (Takaoka et al. 2003), which signals a potential viral infection and hence a quite different problem, will be interesting to test.

Importantly, it has recently been claimed that potentially the whole TP53 regulatory region for the major transcript undergoes extensive chromatin remodeling in different stages of the cell cycle, effectively becoming "open" when the cell goes into G1 phase, and "closed" when the cell is in G0 phase (Su et al., 2009). Further, the shared regulatory region of TP53 and WRAP53 includes the recognizable remains of a LINE element in all placental mammals, about 500 bp upstream of the main TP53 transcription start site (TSS). Slightly downstream of this, the mouse and rat show the additional insertion of a SINE (UCSC browser, Kent et al. 2002, http://genome.ucsc.edu/). It will be important to understanding the regulation of these most important of all cancer related genes, to understand nucleosome occupancy under different conditions, particularly stages of the cell cycle (e.g., Hogan et al. 2006).

## 2 Materials and Methods

Our data are 36 bp reads obtained with SOLEXA sequencing. Following Giresi et al. (2007), cells are grown, then after removing media they are immediately treated with a formalin solution and left for ~20 minutes before quenching this reaction with glycine. The cells are collected and the DNA extracted. The DNA is sheared by sonication (so the densest part of the smear is ~100-200bp in size) then phenol chloroform extracted. Typically 1-5% of the total DNA appears on the aqueous (phenol) phase and this is extracted and run on a gel. Fragments ~100 to 200 bp in length are selected and prepared for sequencing on one end using either the SOLEXA genomic DNA kit, or its small sample equivalent that uses extra cycles of PCR (the CHiP-seq protocol). To date results on the procedure indicate that the recovered or "open" regions tend to be those regions of the genome where nucleosomes are not in close contact with the DNA (Giresi et al., 2007). The data we have run are a combination of fraction of the human cell line data provided by Jason Lieb, part of the ENCODE project, and some test sequence image data from horse fibroblasts (Waddell, Xie, Creek, 2008, unpublished).

The bases of the sequences at each position were called with the BUSTARD image analysis software by Illumina. Matching the optimal base calls is a probability file with a probability score for A, C, G, and T at each position for each sequence. The score is rounded to an integer that runs from -40 to +40. This score equals $10\log_{10}(p/(1-p))$, where $p$ is the probability error of that base call. Rearranging this we have the probability that the base is really nucleotide $I$ given as, $p(i) = 10^{z_i}/(1+10^{z_i})$, where $z_i$ equals the Illumina score for base $i$ at that position divided by 10. For example, if the scores are -9 9 -40 -40 then the Illumina score is saying that $p(A) = 10^{-0.9} / (1 + 10^{-0.9}) \sim 0.112$, $p(C) = 10^{0.9} / (1 + 10^{0.9}) \sim 0.888$, $p(G) \sim 10^{-4}$ and $p(T) \sim 10^{-4}$.

We map all our reads to the parent genome using the program LAST (Frith 2009- http://last.cbrc.jp/, Frith et al. 2009). LAST uses a similar alignment algorithm to BLAST, but it also automatically selects the optimal word length to use. It was necessary to use algorithmic alignment methods in this case since the genomes are large ($>10^9$) and the number of reads to be mapped is also large ($>10^7$), making exact methods of finding the optimal alignment (such as dynamic programming, Smith and Waterman 1981) impractical.

With LAST sequences are placed into a FASTA file separated with ">" then they will be processed consecutively. LAST will find all optimal gapped subalignments to the genome with fewer than a specified number of mismatches (e.g., no more than six mismatches). The run time increases with the minimal mismatch distance. Indels due to the SOLEXA sequencing procedure are rare and almost certainly only one base pair in length (as long as the machine is run within recommended specifications).

During the development of this research Martin Frith kindly extended LAST to align bases using the *.PRB probability files (Frith et al., 2009). In addition, LAST sets a penalty for an overriding mismatch. This is set to approximately the expected value of a sequencing error or SNP in the parent genome. If the score is better using this penalty than that from the site specific *.PRB file, then effectively the program is saying that in that position it is more likely to be an difference from the parent genome, rather than a sequencing error, and it thresholds the penalty for this. The latest version of LAST is available at http://last.cbrc.jp/.

## 3 Results
### 3.1 Outline of 4 methods of mapping short reads
Method 1. Unique Matches Only

Unique or unambiguous matches are those, which map to only one region in the target sequences of interest. This is usually set with an upper limit on the total number of nucleotide mismatches (for the Illumina software this has been 2, with LAST it can be arbitrarily high). With FAIRE data, short reads are mapped, and their frequency in a window of fixed size is reported. The window can be slid by regular increments across the genome and reads may also be prorated by fraction of the read that falls in the window. This approach completely discards data reads that map to multiple target regions.

Method 2. Equally Shared Proportions

Here the unambiguous matches are mapped as above. Next reads with two or more equally good mismatch scores to the target (often with a cap of 2 mismatches) are given equal shares. For example, if *n* is the number of equally good mapping locations, each one gets 1/*n* of a read allocated. This method takes account of ambiguous reads, which make up most of the data but assumes any read has equal likelihood of coming from every region to which it can be mapped. This takes no account of the proportion of reads that map in a window when mapping ambiguous reads.

Method 3. Iterated proportions via an Expectation-Maximization (EM) Algorithm

This technique combines elements of the previous methods but uses a simple formula for weighting the contribution of ambiguously mapped reads to a region in proportion to the number of reads already mapping to a window, then iterating.

Thus instead of assuming all ambiguous reads have an equal probability of coming from each region they match to, their contributions to each of these regions are proportioned according to the relative frequency of matches to that region. On each cycle all reads are mapped again with reference to the previous cycle. Once this is done, the values on that completed cycle contribute the prior frequencies for the next cycle. The advantage of this technique is that it takes account of the fact that an ambiguous read has greater empirical likelihood of being from a region of the genome that, under the same conditions, already has more reads mapping to it.

In some situations, "windows" (potential locations) are naturally discrete, as they, for example, in the case of aligned sequence site patterns (Waddell 2005). With FAIRE data, since the positions of nucleosomes are typically unknown in advance and/or dynamic, it is desirable to use a sliding window. In this case, it is the frequency of the window in the prior iteration, which the read is closest to the center of that, is used. It is possible to allocate the whole read to a window or not based on whether more than 50% of it is in the window, and breaking ties arbitrarily (e.g., by putting such cases in the following window). It seems preferable to prorate the reads at the edges of windows rather than allocate them fully or not at all, to a window, and that is the approach used in the examples herein.

Method 4. Probabilistic Iterated Proportions via EM algorithm

Our final method of analysis is an extension of the previous method which factors in the relative probability of the read being a match to a region on the genome based on all information at hand, independent of the frequency of reads mapping to that window. In our case, these mapping probabilities include the probabilities of erroneous base calls coming from the machine, an estimate of the joint indel rate due to either the machine adding or subtracting bases or real indels in the target genome, or differences from the target genome (either due to errors in its sequencing or SNPs and/or mutations). It is usually appropriate to assume that a read comes from the target genome (especially in controlled experiments), but still have a mismatch (or probability) threshold below which reads (and their error estimates) are discarded as questionable in either quality and/or source.

**3.2 Further details of the EM methods**

A simple EM algorithm to map these data ignoring any mismatches is now described in further detail. It is like the EM algorithm of Waddell (2005) used to map ambiguous site patterns in a sequence alignment. In that application, if the state of only one species for the site pattern is ambiguous, then that site pattern can be matched in theory to four unambiguous patterns. The ambiguous site pattern is thus a subset of the information in the complete site patterns with which it is compatible. In the current situation, we can imagine the multiply mapped reads (analogous to ambiguous site patterns) to be substrings (or probable substrings) of larger strings (fully resolved site patterns) found at FAIRE sites.

First, map all the unambiguously mapped sites to the genome. Next select an analysis window width. A typical window width for FAIRE is ~200 bp. The window is slid over the genome *x* bases at a time. The score of the window is equal to the number of unambiguously mapped reads within it (with partial overlaps of reads prorated). For the moment we will assume all reads are uniformly distributed within a FAIRE region; this assumption may be revisited in future. For each ambiguous read, consider the total score (on the previous cycle) of the window it is closest to the middle of, for all *k* positions of the genome it maps to. Let these *k* scores be $\bar{x}(w_j)$. Now assign the ambiguous read to be proportional to these location window values. That is, the estimated frequency of the *i*-th location of an ambiguous read is,

$$f_i^* = \bar{x}(w_i) / \sum_{j=1}^{k} \bar{x}(w_j) \qquad (1)$$

This is the expectation step. Repeat this for all ambiguous reads. The result is the same, irrespective of the order the ambiguous reads are processed in, but each ambiguous read must be reweighted once each cycle, using the values of the window in the previous cycle as the proportionality weights. Continue until the process converges. Since we are primarily interested in a gross level of sequence coverage at different parts of the genome, it is possible to be lax in this criterion, for example stopping when all windows change their values by less than 0.1 from the previous cycle.

To map iteratively along with the probability (independent of the observed frequencies across the genome) of a read mapping to a certain location we use the following equation to allocate reads,

$$f_i^* = p_i \times \bar{x}(w_i) / \sum_{j=1}^{k} p_j \times \bar{x}(w_j) \qquad (2)$$

Here $p_i$ is the probability that a read maps to that location in the genome based on information prior to seeing the frequency of reads at specific locations in the genome.

There are choices in how $p_i$ is generated and interpreted. Usually we will assume that the read comes from DNA from the target genome, and all the probabilities for mapping to different locations will be adjusted to sum to one. To keep the numbers manageable with hopefully a minimal loss of information, we usually discard any read mapping with a probability 10,000 times less than that of the most probable mapped position, and we discard all reads with more than six mismatches (a probable error rate of over 1 in 6).

For the analyses below, we also set the probability of a SNP and/or sequencing error in the parental genome to 1 in 2500. Further, we set the probability of an indel to be 1 in 10,000, with an extension cost of the same.

The starting point can have an effect. Potential methods for the starting point include the uniquely mapping window scores, the equally shared window scores or all windows having an equal score. The last two methods are the same except the former is advanced one iterative cycle. The first method carries the risk of having zero frequency on the first cycle for a window whose only reads are ambiguous, which can then not easily increase. As we see below, the desired accuracy of convergence is tempered by long computational times and some evidence that the regions of the genome that are slowest to converge, may not be of primary interest.

**3.3 Computational Aspects**

Here we illustrate some of the actual computational costs associated with use of our new methods. The EM algorithms require storage of a number of large vectors. These are ideally kept in memory, in order to maximize the speed of the calculations, since access to some of them tends to be randomized with respect to ordering of the reads. Method 3, the basic EM algorithm, will store a vector of read index, and a list of all locations in the genome where it maps. It is practical to put an upper limit on this number, for example, 10 equally good mapping locations (and discard others at the initialization stage). If we assume that each of *x* reads maps to 5 locations on average, then we need to store ~$5x$ integers. For probabilistic mapping, the storage requirements are a bit more demanding, since along with each location, the probability (or relative probability of mapping there) needs to be stored also (for example, as a single precision float, or even the rather coarse ASCI encoded scoring used in the *.PRB files).

Note, that this vector is usually not too memory intensive, but if it is, since it is always accessed in the same order on each cycle, it can often be stored on hard disk (with a buffer in RAM). This has minimal impact as long as the total time to read it is less than the time for a single cycle of the EM algorithm (for example, a SATA disk can load about 100MB per second, so if the total vector is stored in 4GB, and the cycle time is 5 minutes, there should not be an appreciable penalty in speed).

Also needing to be stored are the scores (e.g., a single floating number) of each window on both the current cycle, the previous cycle, and the initial assignments of unambiguous reads. If the genome is $3 \times 10^9$ GB, then the storage requirement is the former number divided by the window offset, multiplied by 8bytes per floating number. So the total storage of these should not exceed, for example, $3 \times 10^9/10 \times 8 \times 3$, or 7.2GB.

Alternative implementations are possible to increase speed at the expense of memory. The script being used here stores windows in an associative array with the score of every single window affected by every single ambiguous read being stored. The windows are being referred to from the read hit location with the help of a hash table. This speeds up the operations but it consumes memory (in the examples herein, over 10 times the memory).

Due to the current implementation for speed, when analyzing human data it is necessary to map chromosome-wise. This mapping is unbiased if the total number of reads being mapped back to the chromosome of interest is proportionately adjusted. That is, if a read hits at 10 equally good places, only 3 of which are on the chromosome of interest, then that read counts as just 3/10 of a read in subsequent analyses. While this ensures the method is unbiased across the chromosome, it is not optimally informative for the EM methods. For example, if a read maps to two places in the genome on different chromosomes, it will receive ½ weight in the initial cycle and then be treated as fixed. It is unknown if the other location in the genome has a higher or lower density of reads.

Part of the time associated with the EM algorithms is the usual mapping required for methods 1 and 2. Table 1 gives examples of these using the program LAST to map a subset of sequences (one lane worth) for the ENCODE data. All sequences are run against human genome build 19 (hg19 at the UCSC browser, Kent et al. 2002) and the table shows that the setup time was nearly constant. Mapping the 12 to 16 million reads using mismatches and an affine gap penalty took 13 to 17 hours. This aspect of the run time, as expected, tends to increase in a near linear fashion with the number of reads to be mapped. Note, in all these instances LAST is not running in parallel, and each job is using only one core of the Intel i7 at a time. Memory usage during this set up phase with LAST was moderate at around 7.2GB with mismatch and about 11 gigabytes for probabilistic mapping.

Switching LAST to probabilistic mapping slightly increased the time to map, in these cases taking from 18 to 26 hours. The mismatch score allowing for sequencing errors, SNPs etc. followed the HOXD70 score matrix (Chiaromonte et al. 2002). The table also gives examples of the convergence of the EM algorithms. In all cases the convergence criterion was no window changed by more than 0.1 from the previous cycle. Here it is clear that mapping sequence probabilities, rather than mismatches, has increased the speed of convergence considerably.

With the standard EM of method 3, roughly ½ million reads per 12-16 million reads map to chromosome 17 with an equally good mismatch score (after reads mapping to more than 10 locations are removed). The current code, optimized for speed at the expense of memory usage, then required about 10GB to run. Switching to method 4, or probabilisitic EM, the total memory requirement remained similar.

Table 1. Details of EM algorithms mapping FAIRE 36bp reads to hg19: Chromosome 17.

| Cell Line | Reads[a] | Database[b] | Mistmatch[c] | EM50[d] | EM100 | EM200 | Probability[c] | EM50 | EM100 | EM200 |
|---|---|---|---|---|---|---|---|---|---|---|
| GM12878 | 11.9M | 0:28 | 12h32m | 136 | 323 | 155 | 15h31m | 33 | 39 | 30 |
| HepG2 | 16.0M | 0:28 | 16h45m | 183 | 381 | 158 | 25h39m | 36 | 45 | 44 |
| K562 | 13.6M | 0:28 | 13h27m | 236 | 96 | 345 | 17h31m | 50 | 38 | 25 |

[a] Millions of 36 base pair reads
[b] Minutes to set up the LAST database
[c] Minutes to have LAST map all reads with up to six unweighted mismatches
[d] Cycles of EM required to converge with different window sizes; the first set are mismatches only, the second set are probabilistic

It is also interesting to monitor which windows are converging and which are not. We allowed the iterations to continue much longer than usual and then checked on what was occurring. As figure 1 shows, there has been little change in the support for windows located away from the centromeric region of the chromosome (approximately 22,0000 to 25,0000). Indeed only one other location, near 45,619,000, in the first intron of the major transcript of gene NPEPPS, showed change of more than 0.5 in support. Centromeres tend to contain many repeats. These repeats in turn, have made centromeres difficult to sequence accurately. In addition, centromeres are gene poor and are not usually considered regions of primary interest for the mapping of open regions, which are expected to be associated with transcription start and regulatory sites. The UCSC browser maps virtually no genes in the chromosome 17 centromere.

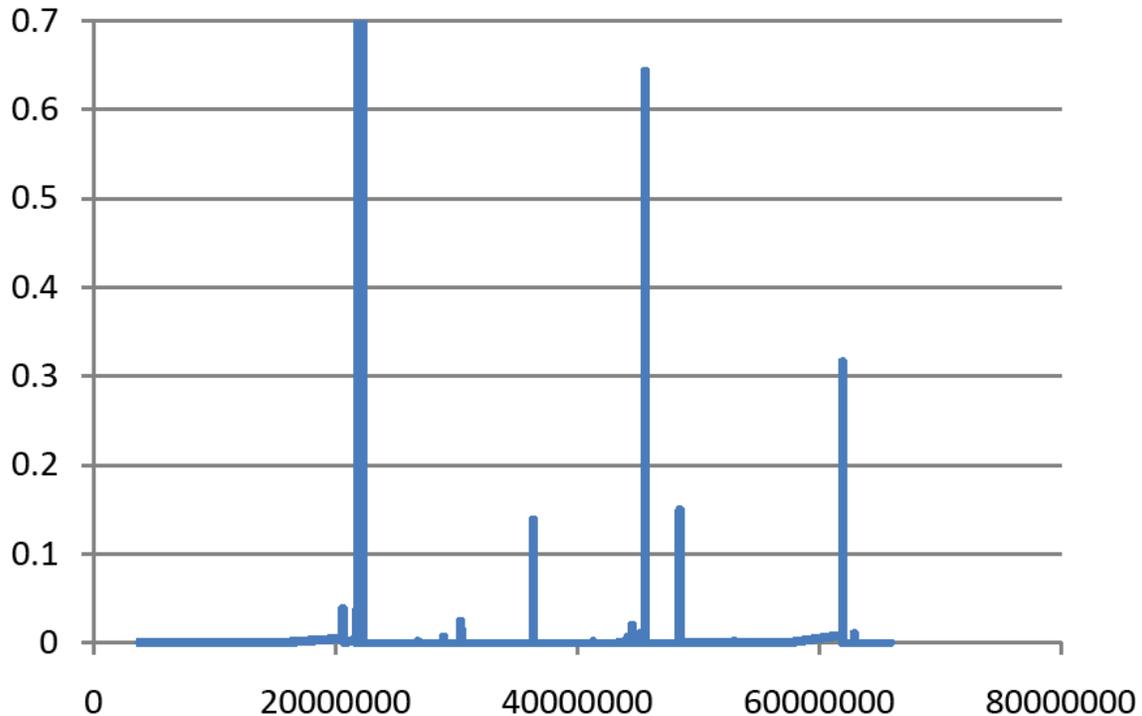

Figure 1. Plot of the location on chromosome 17 versus change in a windows value between cycle 150 and 1050 for probabilistic EM mapping on cell line GM12878. All big changes are positive, while many windows show a miniscule decrease. Regions of the chromosome showing an absolute change of less than 0.001 do not have a blue line. Note a total of 251 windows differed by more than 0.5, of these 7 contiguous windows comprise the spike near 45,000,000.

**3.3 Plots of output for the TP53/WRAP53 region**

Next is an example of methods 1-4 applied to the TP53-WRAP53 region of chromosome 17 in human cell lines being sequenced as part of the ENCODE project. Figure 2 shows that there is indeed a higher density of reads than average near the transcriptional start sites of these genes, although it seems to vary between cell lines. The extent of chromatin remodeling in actively growing cells may also extend quite some way from the transcription start site (Su et al. 2009). Further, looking more closely at the region that contains the major transcriptional start sites of these two genes, then the peaks in this region are about as high as any for the first two cell lines. For the third cell line the general pattern seems to change somewhat. It is interesting to note that TP53 uses a second set of transcriptional start sites close to 7,580,000, which is very close to the highest peak in the third cell line. There is likewise a rarer transcript of WRAP53 that starts close to 7,604,000, which is again very close to the second highest peak in this cell line. It will need to be confirmed with replicates and direct experiments that this cell line indeed, has key nucleosome positions near these two sets of secondary transcription start sites more open than in the other lines.

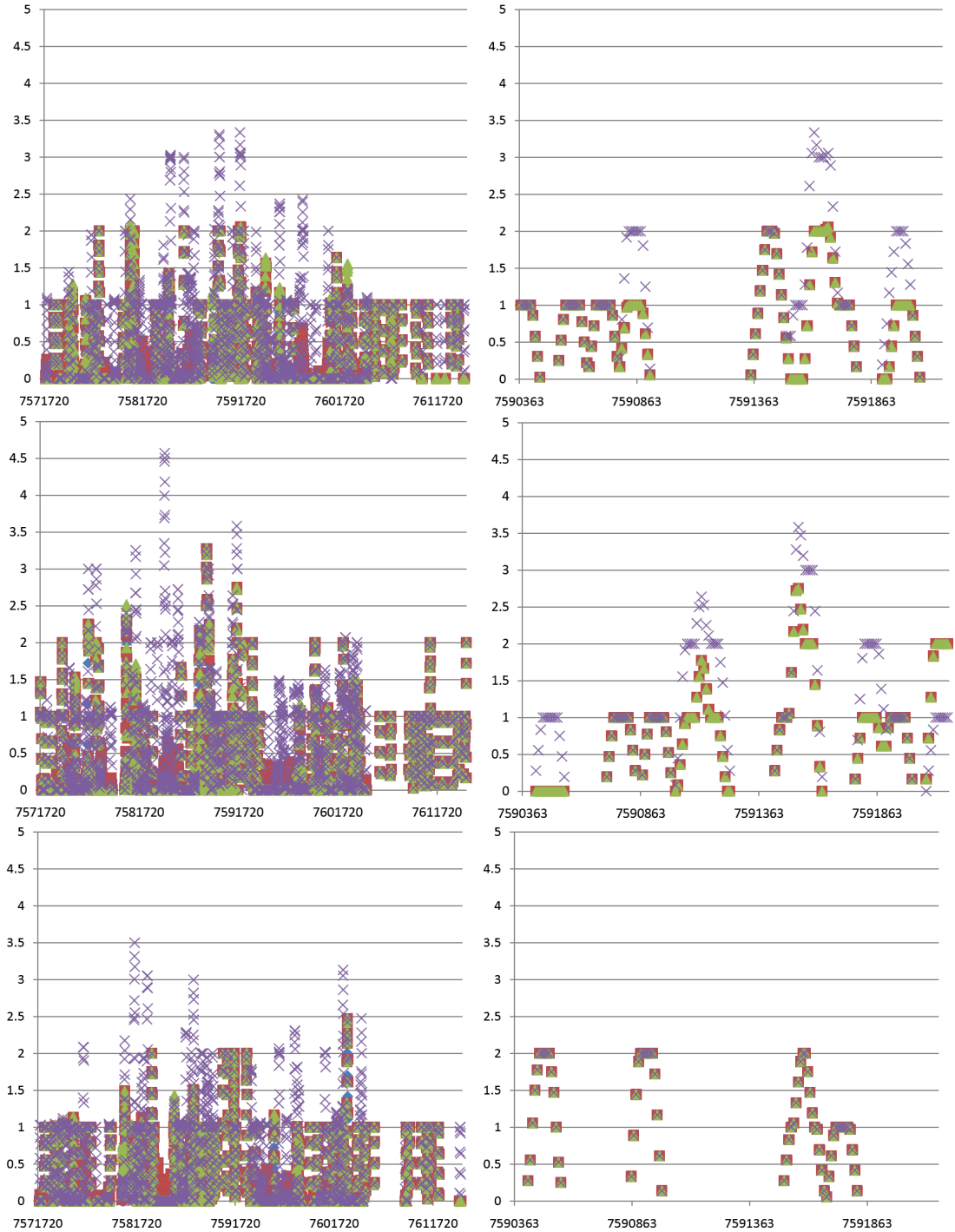

Figure 2. Plots of read density across 100 base pair windows in the region of the TP53 and WRAP53 genes on chromosome 17 according to four methods. The top left figure is a plot for the whole region (genomic coordinates on the *x*-axis for assembly hg19) from the cancer derived cell line GM12878, with read density on the *y*-axis. As expected, the read density is highest nearest the middle of the graph, near where TP53 and WRAP53 have their main head-to-head transcriptional start sites. Method 1 is shown as blue diamonds, method 2 as red squares, method 3 as green triangles and method 4 as purple crosses. The figure on the top right enlarges the region containing the main transcription start site for TP53 at position 7590863 and for WRAP53 at position 7591667. There seem to be peaks in the region of both these start sites. The second row of figures is for the cell line HepG2, and the third row is for the cell line K562.

In these examples, method 4, which is EM with probabilistic mapping, tends to be accentuating the peaks in these specific regions. In this regard it is interesting to note that between the head-to-head start sites of TP53 there is part of a large L1 element, while both the secondary start sites map to the ends of very large introns, which show evidence of many transposable element insertions. These coincide predominantly with the regions where method 4 is giving higher signal than the other methods.

Wider windows also offer the opportunity to capture more information, at the expense of less precision. Figure 3 shows the result of widening the window to 200 base pairs. In this case it has accentuated the major TP53 transcription start site over all others, which might be expected as it is usually the source of the most frequent transcript in this region. As the right hand side of the figure suggests though, there might be some fine scale loss of resolution. In the future this type of data analysis may benefit from combinations of a variety of window sizes, with the upper limit for best detection of unoccupied nucleosome positions potentially being dependent on how many nucleosomes are expected to be absent in a row.

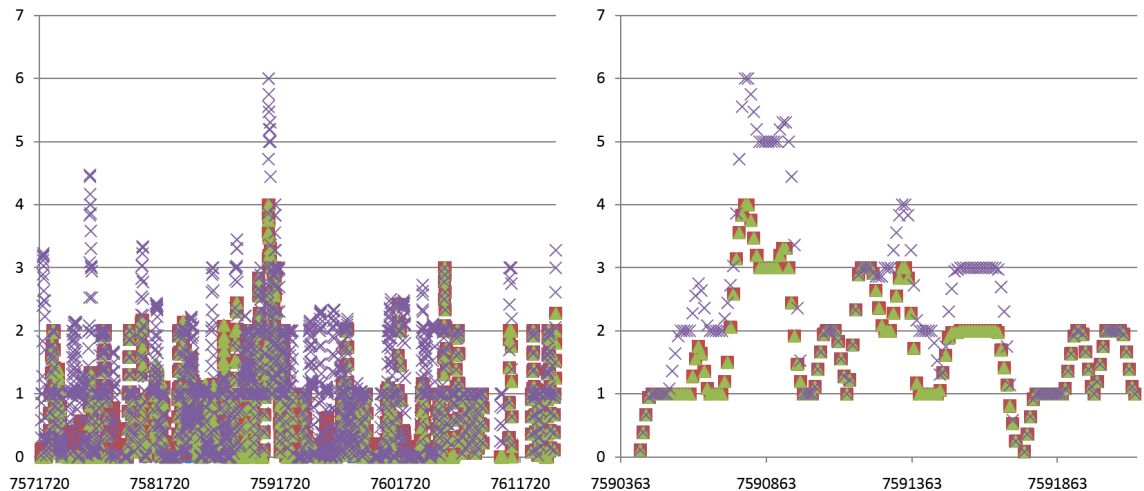

Figure 3. As for figure 2, but using a 200 bp wide window for cell line GM12878.

## 4 Discussion

The use of EM algorithms for mapping large numbers (tens to hundreds of millions) of short (e.g., 36 bp) reads to large genomes with complex repetitive structure (e.g., the human genome) is a powerful statistical method that introduces significant computational challenges. Equations 1 and 2 of this article have strong intuitive appeal, and extend themselves to a wide variety of applications. During the final revisions of this manuscript one of the authors visited Terry Speed, who pointed to the work of Taub et al. (2010), based on similar key ideas. Their results are also bode well for the wide spread use of such methods to better map short reads.

As in many bioinformatics applications, computational issues often come to the fore. While RAM memory usage can be capped by storing the support for the number of windows closest to the center of probable mapping locations of reads, this limits computational speed which can make use of a hash table to quickly locate and update the windows during each iteration. Since this type of data can require a large number of iterations to converge (e.g., hundreds or more) and the total time can run to days, then there is a clear computational trade off. While the distribution of mapping locations for reads tends to be randomized across the genome, it may be useful to look at whether the ordering the mapping of reads can introduce any significant speed increases for this type of data.

Fortunately, the iterative steps are parallelizable. For example, the set of ambiguous reads to be remapped can be broken into sets, mapped separately with reference to the previous full vector, and the results added back together. Finding the optimal way of parallelizing with respect to memory management will be important, since the requirements for random access for each distributed part of the calculations to the previous cycles scores will remain a dominant issue. Another potentially useful avenue will be the adaptation of the algorithm to graphics processing units, which can greatly assist in accelerating some bioinformatics algorithms. However, it is the random access nature of the frequencies from the previous cycle that seems a difficult issue for achieving high throughput.

The EM algorithm is not guaranteed to converge to a global optimum (Dempster et al. 1977). Future work should consider how serious of a problem this can be on a wide range of genome types. The total likelihood of the data is proportional to the product over all reads of the sum over all $k$ mapped locations of the product of the prior probability, $p_i$, multiplied by the assigned empirical weight of the genomic location. If EM not converging to a global optimum it is found to be a problem, then there are variety of techniques to be tried. These include random starts, simulated annealing (Metropolis et al. 1953) and Markov Chain Monte Carlo Methods (Berg 2004). The critical question remains at what computational load? On the positive side, when starting EM from the proportional method 2, which is the current favored approach (e.g., Li et al. 2008), the likelihood will never be lower.

One piece of prior information for mapping short reads that is often not available, but which could prove invaluable, are the background frequency biases introduced by different library preparation/sequencing methods. These biases have recently been shown to be greater than 100 fold per base sequenced (Harismendy et al., 2009). It would also be expected that if they are due to sequence characteristics, they would could be much higher in RNA derived sequences. Possible causes include secondary structure strongly influencing PCR amplification efficiency. It would be highly desirable that the producers of mass sequencing machines publish carefully controlled comparative experiments to allow the extent of these issues to be more accurately gauged. However, with potentially sensitive steps in the sample preparations, the final biases may have a considerable user dependent component, although Harismendy et al. (2009) argue otherwise.

Mapping of support for nucleosome positions (Mavrich et al. 2007, Kaplan et al. 2009) is another potential application of our methods, in what is a highly contentious subject (e.g., Stein et al. 2009). At present, most analyses tend to be discarding ambiguously mapping reads, with the obvious bias against locating nucleosomes around repetitive sequence. Unlike the sonication used with FAIRE, the process of chromatin digestion, when done well, tends to see reads mapping to within a few bases of either end of a nucleosome. For the purpose of centering the diad (middle) of the nucleosome, paired end reads could offer some strong advantages in such an application. Their use may introduce new information to be integrated with the basic EM approach.

## Acknowledgements

This work was supported by NIH grant 5R01LM008626 to PJW and an auxiliary summer supplement training grant supporting TH. Thanks to Jason Leib for access to part of the ENCODE FAIRE data. Thanks to Martin Frith, Sunghwan Yoo, Ariful Azid, Cory Weaver and John Anderson for helpful comments and/or discussions. Special thanks to Marin Frith for implementing our suggestion of probabilistic mapping using *.PRB files into LAST. Shasank Bujimal implemented the initial test version of these methods in PERL, while Cory Weaver reran horse data through the Illumina image analysis pipeline.

## Author contributions

PJW originated the research, developed methods, interpreted analyses, prepared figures and wrote the manuscript. TH checked, modified and ran PERL scripts, produced figures, interpreted analyses and commented on the manuscript.